\documentclass[manuscript]{aastex}

\shorttitle{Stellar occultation by the TNO (229762) 2007 UK$_{126}$}
\shortauthors{Benedetti-Rossi et al.}

\usepackage{multirow}

\usepackage{color}
\definecolor{red}{rgb}{1,0,0}
\definecolor{blue}{rgb}{0,0,1}
\definecolor{black}{rgb}{0,0,0}

\usepackage{ulem}

\begin{document}

\title{Results from the 2014 November 15th multi-chord stellar occultation
by the TNO (229762) 2007 UK$_{126}$}

\author{
G. Benedetti-Rossi$^{1}$, B. Sicardy$^{2}$, M. W. Buie$^{3}$,
J. L. Ortiz$^{4}$, R. Vieira-Martins$^{1,5}$, J. M. Keller$^{6}$,
F. Braga-Ribas$^{7}$, J. I. B. Camargo$^{1,8}$, M. Assafin$^{5}$, N. Morales$^{4}$, R. Duffard$^{4}$, A. Dias-Oliveira$^{1}$, P. Santos-Sanz$^{4}$,  J. Desmars$^{9}$, A. R. Gomes-J\'{u}nior$^{5}$, R. Leiva$^{2,10}$, J. Bardecker$^{11,16}$, J. K. Jr. Bean$^{16}$, A. M. Olsen$^{11}$, D. W. Ruby$^{12,16}$, R. Sumner$^{16}$, A. Thirouin$^{13}$, M. A. G\'{o}mez-Mu\~noz$^{14}$, L. Gutierrez$^{14}$, L. Wasserman$^{13}$, D. Charbonneau$^{15}$, J. Irwin$^{15}$, S. Levine$^{13}$, B. Skiff$^{13}$}

\affil{$^{1}$ Observat\'{o}rio Nacional - ON/MCT\&I, Brazil}

\affil{$^{2}$ LESIA, Observatoire de Paris, CNRS UMR 8109, Universit\'{e} Pierre
et Marie Curie, Universit\'{e} Paris-Diderot, Meudon, France;}

\affil{$^{3}$ Southwest Research Institute, Boulder, CO, USA}

\affil{$^{4}$ Instituto de Astrof\'{i}sica de Andaluc\'{i}a, IAA-CSIC, Apt 3004, 18080 Granada, Spain}

\affil{$^{5}$ Observat\'{o}rio do Valongo - OV/UFRJ, Brazil}

\affil{$^{6}$ California Polytechnic State University, San Luis Obispo, CA, USA}

\affil{$^{7}$ Universidade Tecnol\'{o}gica Federal do Paran\'{a} - UTFPR-DAFIS, Brazil}

\affil{$^{8}$  Laborat\'orio Interinstitucional de e-Astronomia - LIneA, Rio de Janeiro, Brazil}

\affil{$^{9}$ IMCCE, Observatoire de Paris, PSL Research University, CNRS, Sorbonne Universit\'{e}s, UPMC, Univ. Lille 1, 77 Av. Denfert-Rochereau, 75014 Paris, France}

\affil{$^{10}$ Instituto de Astrofísica, Facultad de Física, Pontificia Universidad Católica de Chile, Santiago 7820436, Chile}

\affil{$^{11}$ IOTA, International Occultation Timing Association, USA}

\affil{$^{12}$ University of Nevada, Reno, NV, USA}

\affil{$^{13}$ Lowell Observatory, Flagstaff, AZ, USA}

\affil{$^{14}$ Instituto de Astronom\'{i}a - Universidad Nacional Aut\'{o}noma de
M\'{e}xico - UNAM, Mexico}

\affil{$^{15}$ Harvard-Smithsonian Center for Astrophysics, MA, USA}

\affil{$^{16}$ RECON, Research and Education Collaborative Occultation Network}

\email{aastex-help@aas.org}

\begin{abstract}
We present results derived from the first multi-chord stellar occultation
by the trans-Neptunian object (229762) 2007 UK$_{126}$, observed
on 2014 November 15. The event was observed by the Research and Education Collaborative Occultation Network (RECON) project and International Occultation Timing Association (IOTA) 
collaborators throughout the United States.
Use of two different data analysis methods obtain a satisfactory fit to seven chords, yelding an elliptical fit to the chords with an equatorial radius of $R=338 _{-10} ^{+15}$ km and equivalent radius of $R_{eq}=319_{-7} ^{+14}$ km.
A circular fit also gives a radius of $R=324_{-23} ^{+30}$ km. Assuming that the object is a Maclaurin spheroid with indeterminate aspect angle, and using two published absolute magnitudes for the body, we derive possible ranges
for geometric albedo between $p_{V}=0.159_{-0.013} ^{+0.007}$ and $p_{R}=0.189_{-0.015}^{+0.009}$, and for the body oblateness between $\epsilon=0.105 _{-0.040} ^{+0.050}$ and $\epsilon=0.118 _{-0.048} ^{+0.055}$. For a nominal rotational period of 11.05 h, an upper limit for density of $\rho=1740$ kg~m$^{-3}$ is estimated for the body.
\end{abstract}


\section{INTRODUCTION\label{sec:INTRODUCTION}}

Trans-Neptunian objects (TNOs) are remnants of a collisionally
and dynamically evolved planetesimal disk in the outer solar system.
Their physical characteristics can provide and reveal important clues
about the primordial protoplanetary nebula, planet formation, and
other evolutionary processes \citep{2008AJ....135.1161L}. Moreover,
the inferred chemical, thermal, and collisional processes that they
underwent tell us something about the evolution of the outer Solar
System. However, their large distances make the study
of those bodies difficult, and our knowledge about their sizes, shapes, albedo,
densities, and atmospheres remains fragmentary \citep{2015arXiv151101112P, 2008ssbn.book..161S}.

In the past 25 years, more than 1900 TNOs and Centaurs have been discovered (Minor Planet Center (\citeyear{MPC_LTNO:2015:Online}a,
\citeyear{MPC_LC&SDO:2015:Online}b)). The stellar occultation technique is a very accurate tool to study those bodies, as it provides sizes and shapes at km-level, can detect atmospheres at
nanobar-level (\citet{2011Natur.478..493S, 2012Natur.491..566O}),
and is even sensitive to features such as jets and rings \citep{2015A&A...576A..18O, 2014Natur.508...72B}.
Since 2009, after the first successful observation of a stellar occultation
by a TNO (other than Pluto or Charon) called 2002 TX$_{300}$ \citep{2010Natur.465..897E}, several
objects have been observed by stellar occultations. Examples are Varuna \citep{2010DPS....42.2311S}, Eris \citep{2011Natur.478..493S}, 2003 AZ$_{84}$ (\citealp{2011CBET.2675....1B}, \citeyear{2012DPS....4440201B}),
Quaoar (\citealp{2011AAS...21822412P}, \citealp{2011AAS...21822413S},
\citealp{2013ApJ...773...26B}), Makemake \citep{2012Natur.491..566O},
2002 KX$_{14}$ \citep{2014A&A...571A..48A}, and Centaur objects
like Chariklo \citep{2014Natur.508...72B} and Chiron \citep{2015A&A...576A..18O, 2015Icar..252..271R}.
The observation of a multi-chord stellar occultation on 2014 November 15
increases that list to include the TNO (229762) 2007 UK$_{126}$,
the main topic of this paper.

This TNO was discovered by \citet{2008MPEC....D...38S} in October
2007 with an estimated radius and albedo of $299.5\pm 38.9$ km and $0.167_{-0.038}^{+0.058}$
\citep{2012A&A...541A..92S}, respectively. With a semi-major axis
of 73.81 AU, aphelion distance of 109.7 AU, orbital period of 634.13 yr, orbital eccentricity
of 0.492 and an inclination of 23.34 degrees \citep{JPL_SBDB:2015:Online},
it is usually classified as a Scattered Disk Object - SDO - (according to \citet{2008ssbn.book...43G}) or a member of the class of "Detached Objects" (see eg. \cite{2008AJ....135.1161L}). Moreover, \citet{2011epsc.conf.1078G}
reported the discovery of a companion with a magnitude difference
of 3.79 mag in the F606W band of the Hubble Space Telescope. Its orbit is still unknown, but it is
expected to be non-circular \citep{2014A&A...569A...3T}.

In this paper, we present results derived from the 2014 November 15
stellar occultation by this body. Section
\ref{sec:PREDICTIONS-AND-OBSERVATIONS} briefly describes our prediction
scheme and presents the observations. Data analysis is described in
Section \ref{sec:DATA-ANALYSIS}. The size and shape of the TNO as
well as their physical implications are discussed in Section \ref{sec:SIZE-AND-SHAPE},
before concluding remarks in Section \ref{sec:Conclusion}.

\section{PREDICTIONS AND OBSERVATIONS\label{sec:PREDICTIONS-AND-OBSERVATIONS}}

The 2014 November 15 occultation was identified in a systematic search
for TNO occultation candidate stars, made at the 2.2 m telescope of
ESO, using the Wide Field Imager (WFI). This search yielded local astrometric
catalogs for 5 Centaurs and 34 TNOs (plus Pluto and its moons) up to 2015, and for stars with magnitudes as faint as R $\sim$19. Further details can be found in \citeauthor{2010A&A...515A..32A} (2010, \citeyear{2012A&A...541A.142A}), \citet{2014A&A...561A..37C} and \citet{2015A&A...584A..96D}.

After identifying the target star, astrometric updates of the star
UCAC4 448-006503 (UCAC2 31623811, R=15.7) close in time to the predicted occultation
were performed with the 60 cm telescope at Pico dos Dias Observatory
(OPD/LNA - IAU code 874) and with the 77 cm telescope at La
Hita Observatory (IAU code I95). From OPD, 20 images with
45 s exposure time were acquired using Johnson's I filter (centered at 800 nm)
and an IkonL 9867 CCD camera on October 19, 2014. From La Hita, 99 unfiltered images of 400 s exposure time were obtained on October,
29, 30 and 31, 2014, with the 4k $\times$ 4k camera, which provided a very
wide field of view of 47 $\times$ 47 arcmin. In both cases, the images were
obtained at times when the objects were near the meridian, to minimize
possible Differential Chromatic Refraction (DCR) problems. Unfortunately,
since the apparent visual magnitude of the TNO is approximately 20.1,
the signal to noise of 2007 UK$_{126}$ in the individual exposures
was poor (around 4 to 8, depending mainly on the seeing) and did not allow us to obtain more accurate astrometry of the TNO.

To derive accurate astrometry of 2007 UK$_{126}$, 36 images of
the TNO were obtained with the Calar Alto (IAU code 493) 1.2m telescope
using the 4k $\times$ 4k DLR CCD camera on October 28 and 29, 2014. The camera
provides a field of view of 22 $\times$ 22 arcmin. Exposure times were
400 s, which allowed us to obtain a signal to noise ratio on the target
larger than 40, with no filter. The images were also obtained when the object
was near culmination to minimize any possible problem due to DCR.
The astrometry provided the offsets in right ascension and declination with respect to
the nominal positions based on the JPL Horizons ephemeris.
From the dispersion of the offset measurements (i.e. the quadratic sum of the 1-$\sigma$ uncertainty in the ephemeris update (20 mas) and the 1-$\sigma$ error in the position of the star (4 mas)) the final uncertainty
in the prediction was estimated about 20.4 mas, comparable
to the expected shadow path width. The final prediction indicated
that the shadow was favorable for observers in several states in the USA (Fig. \ref{fig:PredicMap}). A compilation of our measurements provides the following
ICRF/J2000 star position at the date of the occultation:

\begin{eqnarray}
\alpha=+04^{h}29^{m}30.^{s}6100\pm0.''022\label{eq:a_d_star}\\
\delta=-00^{\circ}28'20.''908\pm0.''023\nonumber 
\end{eqnarray}

The RECON project \citep{Buie2016RECON} pilot sites and other potential sites participated
in the campaign for a total of 20 different stations
(Tables \ref{tab:SitesPositive} and \ref{tab:SitesNegative}). Bad weather conditions spoiled observations in 11 sites.
Meanwhile, six sites from RECON, one IOTA (International Occultation Timing Association) site located in Urbana, Illinois, and two telescopes at San Pedro
Martir acquired data, for a total of 7 positive detections of the event
and two negative chords at San Pedro Martir,
which were located more than 500 km south of the shadow path (Fig.
\ref{fig:PredicMap}). The times of the star disappearances (ingress) and re-appearances (egress) for the seven detections are listed on Table \ref{tab:Instant_Ingress_egress}.

All acquired data, with exception of the data from the two telescopes in San
Pedro Martir, were in video format. A video time inserter (VTI) was
used to place a time-stamp on each individual video frame. All RECON sites use IOTA-VTI that The
VTI interacts with a GPS receiver to obtain the time, ideally with an absolute accuracy
of a few milliseconds, and then superimpose the current time on each video field as it passes from the camera to the computer (see \cite{Buie2016RECON}). Unfortunately, no information is saved on any image header because of the video format.

\section{DATA ANALYSIS\label{sec:DATA-ANALYSIS}}

Considerable detail is provided on video data that is applicable to
RECON data in \citet{Buie2016RECON}.  In particular, the discussion
on frame and field interleaving of the data is especially relevant for these
data. All of the RECON data were collected with a SENSEUP value of 128x,
meaning that the integration time is equal to 128 times the field rate of
the NTSC video signal, which resultted in integrations of approximately 2 seconds. The Urbana data were collected with a Watec camera integrating 128 video frames, or approximately 4 seconds. If there are
no dropped frames or other problems, the RECON video will have 64 copies of
each integration in the video data stream. The Urbana video will have 128
copies.

Two independent analyses were performed on the video data to extract
light curves and timing information. The two approaches are different
enough to be useful as a cross-check of results to provide information
on the uncertainties in the final projected shape.  The primary analysis
is from Benedetti-Rossi (GBR) and the secondary analysis is from Buie (MWB).

\subsection{OCCULTATION LIGHT CURVES\label{sub:OCCULTATION-LIGHT-CURVES}}

\subsubsection{GBR extraction\label{GBR_extraction}}

Using \citeauthor{Audela:2015:Online} (a free and open source astronomy software for digital observations: CCD cameras, Webcams, etc.) we extracted individual frames to FITS format from the video at a rate of 29.97 frames per second. A careful check was done in all sets of images to verify whether the extracted time corresponds to the time printed at each frame.

The frames were then grouped to match the corresponding SENSEUP value. The first and last frames of each 64-frame (or 128) sequence were identified by counting frames from a calibrated starting point and checked with a change in field brightness. They were then excluded and the other 62 (or 126) frames were averaged to obtain each image that corresponds to an individual exposure time. By not considering the first and last frame from a sequence, we avoid the need to de-interlace and re-interlace the frames, and we do not take into account any field from the previous or next sequence. Note that this process also preserves the mid time of each image which was extracted from the mid-frame of each sequence. This method generates times with a systematic shift with respect to the absolute times (one integration cycle) but does not affect the final shape when all chords are processed the same way. This whole process of converting video to separate stack image files requires special attention because of possible dropped frames, duplicated fields, or any incompatibility with different software packages, drivers, or plug-ins.

Differential aperture photometry was extracted from the data using the PRAIA package \citep{2011gfun.conf...85A} to obtain the light curves. Two field stars with 8 pixel and a third one with 10 pixel photometric aperture were used to calibrate the occulted star flux. Sky background flux was obtained from an annulus of internal radius of 16 and external radius of 20 pixels around the first two calibration stars and 20 to 24 pixels around the third one. For the occulted star, a photometric aperture of 7 pixels was used while for the sky background flux an annulus of 10 pixels inner radius and 16 pixels for the outer radius was used.

The occulted star flux was then normalized to the unocculted stellar flux by applying a third-degree polynomial fit to the flux just before and after the event. The resulting light curves are shown in Figure \ref{fig:LightCurves}.

\subsubsection{MWB extraction\label{MWB_extraction}}

Accurate determination of the time of each integration is discussed
in detail in \cite{Buie2016RECON} and requires locating the exact place
in the video stream where a new integration is first seen.

In this analysis, custom software was created in IDL for each site's data.
The first step requires converting the AVI-format video data to individual
image files. The freely available tool, \textit{ffmpeg}\footnote{https://www.ffmpeg.org/}, was used to extract a
sequence of images to individual PNG format files. In all cases only 90-120
seconds of data were extracted, centered if possible on the occultation
chord.

The basic flow of data processing contains some or all of the following
steps, customized for each dataset. 1) Create a mean sky image. If
possible to build, this image contains the general background gradient, typically from
amplifier glow in one corner as well as hot pixels. Sites where the
tracking was perfect could not be corrected since separate dark or sky images
were not collected. Mean sky images require a rather complicated stacking
of frames. The first step is to perform a robust stack of one image per
integration. There are as many of these stacks as there are frames in
an integration and these stacks are also robustly averaged into the final
mean sky image. If the entire cube is stacked at once, the frame replication
for the integrations will subvert the robust averaging algorithm.
2) Subtract mean sky image from each frame. 3) Re-interlace the images, if
needed. The need for re-interlacing is easily seen in a raw light curve of
a bright star. Each integration has a unique signal level and will look like
steps from one integration to the next. If the signal steps cleanly between
integrations, the interlacing is correct. If there is a single frame point
between the two levels, re-interlacing is required. 4) Extract source and comparison
star fluxes from each frame. 5) Down-sample by averaging to a single
measurement for each integration. 6) Apply timing formula from \citet{Buie2016RECON} to get absolute timing of each data point. The deviations from
this set of steps is now described for each site in turn.

The photometric extraction required some special handling in all cases.
Four nearby field stars were measured to obtain both position and flux.
When the occultation star was visible, its position was also measured.
The offset for the occultation star relative to the brightest field star
was determined for each dataset. This offset was used as the exact position
for extracting the occultation star flux on each image. This process
avoids aperture wander off of the occultation star location during the
occultation. Each site reduction process is presented as follows.

{\em Ruby/Reno} -- A mean sky frame was generated and subtracted.  The data were re-interlaced and had no dropped frames.  The photometry was generated from a 5-pixel object aperture and a sky annulus from 8-40 pixels.  The raw photometry shows clear signs of degrading sky conditions from 10:18 -- 10:20 UT.  Getting the timing required a more careful examination of the images and determination of the brightness
of the transitional frames for the model timing extractions.

{\em Sumner/Carson City (S)} -- A sky frame was generated and subtracted.
The images were re-interlaced and had the same dropped frame problem
found with the other Carson City site. The final results have
full-quality data after cleanup. The photometry was generated with a
5-pixel object aperture and the sky was determined from a robust mean of
the entire image which was flat due to the subtracted sky mean.

{\em Jack C. Davis Observatory/Carson City (B)} -- The mean sky image could not
be generated to high-quality track and no calibration images. The data
required re-interlacing and also suffered from dropped frames. Each frame
was manually inspected to identify corrupted frames (one or two every 3
seconds). During the manual inspection the IOTA-VTI timing information was
used to establish the identity of each frame and it's association with the
individual integrations. In some cases, all 64 frames were good, but many
had one or two frames that were not used. When properly identified, the
timing is not affected by the loss of a few frames out of the 64 copies that
should have been collected. In no case was there a dropped frame on
an integration boundary. The photometric used was a 5-pixel object aperture
and a sky annulus of 8-25 pixels. The final result, though laborious, was
not affected by the dropped frames.

{\em Yerington} -- A mean sky frame was generated and subtracted.
The data were re-interlaced and had no dropped frames. The photometry
was generated from a 5-pixel object aperture and sky was determined from the
mean of each frame.

{\em Bardecker/Gardnerville} -- The mean sky image could not be generated
due to the excellent tracking at this site and the lack of separate calibration
image sequences.  The video data required re-interlacing but did not have
any instances of dropped frames in the video data stream. A 5-pixel photometric
aperture was used for the sources and a local sky value was determined
for each with a sky annulus of 8--40 pixels. The frame integration
boundary was visually determined by seeing the change in the background noise.
A single transition was sufficient to establish the timing for the entire
sequence. Note that this site used a telescope with an equatorial mount.

{\em Tonopah} -- A mean sky frame was generated and subtracted.
The data were re-interlaced and had no dropped frames. The photometry
was generated from a 5-pixel object aperture and sky was determined from the
mean of each frame.

{\em Olsen/Urbana} -- This dataset uses a similar but not identical setup
to the standard RECON system. The camera was a Watec-120N+ and has a similar
sensitivity and operation to the RECON MallinCAM cameras but can integrate
twice as long.  The timing was provided by a Kiwi OSD video time inserter.
A sky frame was generated and subtracted. The images were re-interelaced
and there were no dropped frames. A 5-pixel photometric aperture was used
with a sky annulus of 8-50 pixels to remove the small amount of flat
sky residual background.

All the resulting light curves are shown in Figure \ref{fig:LightCurves}.

\subsection{OCCULTATION TIMING\label{sub:OCCULTATION-TIMING}}

As described by \citet{2013ApJ...773...26B}, the start and end times
of the occultation were obtained for each light curve by fitting a
sharp edge occultation model. This model is convolved by Fresnel diffraction,
the CCD bandwidth, the stellar apparent angular diameter in kilometers, and the finite
integration time (see \citet{2009Icar..199..458W} for more details).

The Fresnel scale ($F=\sqrt{\lambda D/2}$ ) for the geocentric distance
D = 42.6 AU (or $6.37\times10^{9}$ km) of 2007 $UK_{126}$ by the time
of the event is approximately 1.4 km for a typical wavelength of $\lambda=0.65$
$\mu$m. The star apparent angular diameter is estimated using the formulae of \citet{1999PASP..111.1515V}.
Its B, V , and K apparent magnitudes are 16.2, 15.6, and 13.7, respectively,
in the NOMAD catalog \citep{2004AAS...205.4815Z}. This yields a diameter
of about 0.3 km projected at the distance of the 2007 $UK_{126}$.
The smallest integration time used in the positive observations was
2 s, which translates to almost 48 km in the celestial plane. Therefore,
the occultation light curves are largely dominated by the integration
times, not by Fresnel diffraction or the star diameter.

The occultation fits consist of minimizing a classical $\chi^{2}$
function for each light curve, as described in \citet{2011Natur.478..493S}.
The free parameter to adjust is the ingress (disappearance) or egress
(re-appearance) time, which provides the minimum value
of $\chi^{2}$ denoted as $\chi_{min}^{2}$. The best fits to the
occultation light curves are shown in Figure \ref{fig:LightCurves},
and the derived instants of ingress
and egress are shown in Table \ref{tab:Instant_Ingress_egress}.

Note in Fig. \ref{fig:LightCurves} that the disappearance and reappearance of the star is very clear with exception of the Reno observation. The Reno data required special care due to the degrading sky conditions after 10:18 UT. At first glance, the point in the light curve near 10:19:32 seems to indicate the start of the occultation. However, careful examination of this integration still shows a faint remnant of the star along with some image artifacts and leads to an anomalously low signal. The integration at 10:19:34 clearly shows the star and is the frame where ingress begins as the star is completely gone by the next integration. In general, one might think the point at 10:19:32 is perhaps indicating some other interesting lightcurve feature other than ingress. In this case, the degrading observing conditions argue that the measured flux from this integration is spurious and should be treated as an un-occulted timestep and that is what we adopt for the interpretation of those light curves.

\subsection{LIMB FITTING\label{sub:LIMB-FITTING}}

Objects with diameter larger than about 1000~km are expected to be in hydrostatic equilibrium. 
As such, they reach either Maclaurin spheroid or Jacobi ellipsoid states \citep{EllipsoidalFig}.
The critical diameter, 
defined as the minimum size necessary  to reach hydrostatic equilibrium, 
can be estimated to 200-900 km for icy bodies or from
500-1200 km for rocky bodies \citep{2008Icar..195..851T}. 
2007 UK$_{126}$ is within those critical ranges and considering it is in
the small angular momentum regime (i.e., the body presents a low rotation period with an estimated rotational period of 11.05 hours and it presents a small-amplitude rotational
light curve with $\Delta m=0.03\pm0.01$ mag \citep{2014A&A...569A...3T}). Thus,  
we will assume here that this TNO is close to the Maclaurin state.

Consequently, its limb is elliptical and is characterized by $M=5$ adjustable parameters: 
the coordinates of the body center, relative to the star in the plane of the sky ($f_{c}$, $g_{c}$); 
the apparent semi-major axis $a'$; the apparent oblateness $\epsilon'=(a'-b')/a'$ 
(where $b'$ is the apparent semi-minor axis);
and the position angle $PA$ of the semi-minor axis $b'$. 
The position angle $PA$ is counted positively from the direction of celestial North to celestial East, 
while the quantities ($f_{c}$, $g_{c}$) are expressed in kilometers, 
positively toward the celestial East and North directions, respectively. 
In the oblate Maclaurin spheroid hypothesis, 
the apparent oblateness $\epsilon'$ is related to the true oblateness $\epsilon=1-(c/a)$
(where $a=a'$ and $c$ are the true equatorial and polar radii, respectively) through:
\begin{equation}
\epsilon'=1-\sqrt{\cos^{2}(\xi)+(1-\epsilon)^{2}\sin^{2}(\xi)},
\label{eq:true_oblat}
\end{equation}
where $\xi$ is the polar aspect angle, i.e. the angle  between the polar $c$-axis and the line of sight.
The case $\xi=0^{\circ}$ (respectively $\xi=90^{\circ}$ ) then corresponds to the pole-on (respectively equator-on) geometry. 

Note that  2007~UK$_{126}$'s pole direction is currently unknown, so $\xi$ is undefined. 
Finally, it is useful to quantify the size of the body through its apparent equivalent radius 
$R_{eq}$ (rather than its equatorial radius $a$) defined by 
$R_{eq}=\sqrt{a'b'}=a'\sqrt{1-\epsilon'}$.
This corresponds to the radius of the disk that has the same area as that enclosed by the apparent limb.

The seven positive occultation chords provide 
$N = 14$ data points (the ingress and egress chord extremities, see Fig.~\ref{fig:NominalSolutionGBR} and Table \ref{tab:Instant_Ingress_egress}),
whose positions are denoted as $f_{i,obs}$, $g_{i,obs}$. 
The best elliptical fit to those points minimizes the radial residuals, 
from which the relevant $\chi^{2}$ function is defined.
The quality of the fit is then assessed through the value of the 
$\chi^{2}$ function per degree of freedom (or unbiased $\chi^{2}$), defined as
$\chi_{pdf}^{2}=\chi^{2}/(N-M)$,  see e.g. \cite{2011Natur.478..493S} for details. 
The 1-$\sigma$ level uncertainty of a given parameter is then obtained by varying it
(all the other parameters being re-adjusted during this exploration) 
so that the $\chi^{2}$ function varies from its minimum value 
$\chi_{min}^{2}$ to $\chi_{min}^{2}+1$.

\section{RESULTS\label{sec:SIZE-AND-SHAPE}}

Figs.~\ref{fig:NominalSolutionGBR} and \ref{fig:NominalSolutionMWB} 
display the best elliptical limb fits obtained from the two sets of timings (GBR and MWB). Note that 
the difference between the immersion and emmersion times comes from the fact that the determination of the instants is very sensitive to the photometry. Since the star is very faint and its flux is close to the background sky flux, a small change on the light curve can induce different instants.

Although having slightly different times, the parameters derived from each method agree with each other at the 1-$\sigma$ level (Table~\ref{tab:Physical-Parameters}).
Moreover, the respective values of $\chi^2_{pdf}$ for the elliptical fits are under 0.7,
indicating satisfactory adjustements of the model to the data.

Purely circular shapes were also fitted to the data, resulting in $\chi_{pdf}^{2}$ values of 
1.69 and 1.46 for the GBR and MWB timings, respectively. 
Thus, although circular fits remain acceptable in terms of quality, they do degrade the value
of $\chi_{pdf}^{2}$. 
In fact, Table~\ref{tab:Physical-Parameters} shows that the 
apparent oblatenesses $\epsilon'$ differ from zero at the $\sim$ 2.5-$\sigma$ level, 
a marginal detection of  non-sphericity for 2007~UK$_{126}$. 

Note that in both solutions the Urbana chord seems a bit displaced towards East 
(Figs.~\ref{fig:NominalSolutionGBR} and \ref{fig:NominalSolutionMWB}),
compared to the elliptic model. 
Since the TNO is near the limit of the critical range to be a Maclaurin object, 
we can consider two possibilities: 
(1) all the timings are correct within 1$\sigma$ uncertainty, implying that 2007 UK$_{126}$ is compatible 
with a Maclaurin object and has large topographic features (craters or mountains) of the order of 10s of km (a possible solution, considering the New Horizons images on Charon that shows topographic features as big as +/- 6 km \cite{2016arXiv160300821N})
or 
(2) 2007 UK$_{126}$ is a smooth (no topographic features) Maclaurin object, implying that timing problems are present 
at some stations. 

With this in mind, we have allowed time shifts for all stations, aligning the middles of all chords,
resulting in a $\chi_{pdf}^{2}$ value of 1.37 and a radial rms of 16 km, i.e. without significant improvement of the fit quality.
Moreover, this implies that all sites had timing issues, some of them as big as one second (half of the integration time), which is unlikely to happen.

The fact that the observations are not repeatable makes difficult the assessment of timing errors.
However, we do not expect large absolute timing errors at the various stations.
Although it is important to note that the process of converting video to the stack images on GBR analysis (Section \ref{GBR_extraction}) may present some small error in time. For the MWB analysis, the conversion from video data to a light curve does not present any intrinsic timing errors beyond limitations imposed by the low SNR for the event (Section \ref{MWB_extraction}).

\subsection{PHYSICAL PROPERTIES FOR 2007 UK$_{126}$\label{sub:PHYSICAL-PROPERTIES-FOR}}

From the size and shape, the density can be derived if the mass of the system is known,
e.g. through the motion of a satelite.
As mentioned in Section~\ref{sec:INTRODUCTION},
\citet{2011epsc.conf.1078G} reported the discovery of a companion,
but no orbital elements are presently available for it. 
However, constraints on the density can still be derived in the Maclaurin hypothesis,
when combined to the rotation period $P$, using the equilibrium equation
\citep{1919MNRAS..80...26P}:
\begin{equation}
\rho = \frac{4 \pi \sin^3(\theta)}
{G  \cos(\theta) \cdot \{2 \theta [2+ \cos(2 \theta)] - 3 \cdot \sin(2 \theta) \} \cdot P^2},
\label{eq:density_maclaurin}
\end{equation}
where G is the gravitational constant (G = 6.67408 $\cdot 10^{-11}$ m$^3$ kg$^{-1}$ s$^{-2}$) and 
$\theta$ is related to the real oblatenes, $\epsilon$, by $\cos(\theta) = 1-\epsilon$.

Eq.~\ref{eq:true_oblat} imposes that the true oblateness $\epsilon$ is bounded according to 
$\epsilon' \leq \epsilon $. On the other hand, Maclaurin spheroids can have a maximum oblateness of $\epsilon = 0.417$. For $\epsilon>0.417$, only triaxial ellipsoid of equilibrium are possible \citep{EllipsoidalFig}.
Moreover, the polar aspect angle must satisfy
$\arcsin(\sqrt{((1-\epsilon')^2-1)/((1-\epsilon)^2-1)}) \leq \xi \leq \pi/2$,
the lower limit case corresponding to  $\epsilon= 0.417$ and the upper
limit case corresponding to  $\epsilon = \epsilon' $. 
In that context, it is interesting to assess the density of probability for $\epsilon$, beyond merely stating
that it should lie in the interval $[\epsilon' , 0.417]$.

For an elliptical fit the orientation of 2007 UK$_{126}$'s pole angle is partially constrained by the ellipse orientation. The polar aspect angle $\xi$ remains undetermined but constrained to an interval [0.45 , $\pi/2$] through Eq. \ref{eq:true_oblat} and constraints on $\epsilon$. Because the position angle of the fitted elliptical limb is known (Table \ref{tab:Physical-Parameters}), we can assume here that the density of probability for $\xi$ is uniformly distributed over all the possible values given by the interval.

Consequently, the probability to have $\xi$ in the interval is $\sim 0.7$, 
considering the possible values of $\epsilon'$ (Table~\ref{tab:Physical-Parameters}).
In other words, the measured apparent oblateness $\epsilon'$ does not require a very specific, 
fine tuned orientation for the aspect angle.

Using both the GBR and MWB solutions we obtain a lowest possible value for $\epsilon'$ (i.e. 0.105 - 0.0040 = 0.065) from wich we can derive an upper limit for the density of of $\rho=1740$ kg~m$^{-3}$

The preferred rotation period for 2007 UK$_{126}$ is P = 11.05 h \citep{2014A&A...569A...3T}. However, other possible aliases exist, and only a secure lower limit of P$>8$ h is eventually given by those authors.
Fig.~\ref{fig:maclaurin_and_density} presents the Maclaurin equilibrium curve for both periods.

To numerically estimate the probability distribution, $P_r(\rho)$, for the density, we generate a sample of aspect angles from an uniform distribution in the valid interval. This in turn provides the density of probability for the density via Eq.~\ref{eq:true_oblat} and Eq.~\ref{eq:density_maclaurin}. We use the lowest 1-sigma value for $\epsilon'= 0.065$ and the prefered value for T=11.05 h.

Fig.~\ref{fig:maclaurin_and_density} also displays $P_r(\rho)$. It shows that although $\rho$ can be in the whole range [320 , 1740] kg~m$^{-3}$ it is probably close to its upper limit, indicating an icy-body. For a lowest rotation period of P=8 h, the probability distribution for the density will be qualitatively the same but in the range [600 , 3300] kg~m$^{-3}$ indicated in Fig.~\ref{fig:maclaurin_and_density} for reference. In this case the density is probably close to its upper limit of 3300 kg~m$^{-3}$ indicating a rocky-body. Only with an accurate rotation period measurement can a more definitive conclusion for the density be stated.

From the equivalent radius $R_{eq}$, we obtain the geometric albedo $p$:
\begin{equation}
p=(AU_{km}/R_{eq})^{2}\times10^{0.4(H_{\odot}-H)} ,\label{eq:albedo}
\end{equation}
where $AU_{km}=1.49598\times10^{8}$ km, $H_{\odot}$ is the Sun magnitude
at 1 AU ($H_{\odot,V}=-26.74$), and $H$ is the object absolute magnitude of the object. Minor Planet Center provides $H_{R}=3.4$ and Perna et al (\citeyear{2010A&A...510A..53P},
\citeyear{2013A&A...554A..49P}) give $H_{V}=3.59\pm0.04$. Adopting the ranges of equivalent radii obtained for both solutions, we calculate the geometric albedo of 2007 UK$_{126}$ in the visible ($p_{V}$ and $p_{R}$). The error bars represent the range
of the albedo obtained for a given solution, combined with the uncertainty
in absolute magnitude. Results are presented in Table \ref{tab:Physical-Parameters}.

\section{Conclusion\label{sec:Conclusion}}

We observed the first multi-chord stellar occultations by the trans-Neptunian
object (229762) 2007 UK$_{126}$. The shadow crossed the United States
on 2014 November 15 and we obtained 7 positive chords, from which we 
obtain the equivalent radius and geometric albedo of the body,
and  an upper limit for its density.

This is the first TNO occultation result from the RECON project. It was observed during the pilot phase with only a few sites in operation. For future occultations, with sufficient astrometric support, five times as many RECON sites will be available for observations.

We present in this paper two independent analyses that provide consistent solutions for 2007 UK126's limb shape. The GBR solution gives an equivalent radius $R_{eq}=319_{-7}^{+14}$ km and geometric albedo that may vary $p_{V} =0.159_{-0.013}^{+0.007}$ to $p_{R} = 0.189_{-0.015}^{+0.009}$, depending on the adopted absolute magnitude. The MWB solution provides an equivalent radius of $R_{eq}=319_{-6}^{+12}$ km and albedo that may vary from
$p_{V} = 0.159 _{-0.011}^{+0.006}$ to $p_{R} = 0.189_{-0.013}^{+0.008}$. The two solutions give comparable minimum $\chi^{2}$ per degree of freedom (0.59 and 0.56, respectively). The equivalent radii we derive here are consistent with, but more accurate than the value based on Herschel observations, R$_{eq}$= 299.5 $\pm$38.5 km \citep{2012A&A...541A..92S}.

A range for density $\rho$ was estimated to be [318 , 1740] kg~m$^{-3}$ using a lowest apparent oblateness from the \textit{GBR} solution and 
considering the rotation period of 11.05 hours. Those values are comparable to other TNOs densities found in literature (as presented in \cite{2012AREPS..40..467B}, \cite{2016PASP..128a8011S} and references). No other information could be derived since there is insufficient orbital information available. 
Obtaining an orbital solution for the satellite of 2007 UK$_{126}$  would be an importnat step foward, 
as it would firmly constrain the mass of the primary, and thus, its density from our size measurement.

The occultation chords also shows that there is no evidence of a very close binary and no satellite could be detected from the shadow track direction.

\acknowledgments

The authors want to thank Barclay Anderson, Charley Arrowsmith, Buck Bateson, Clair Blackburn, Teralyn Blackburn, Mystery Brown, Brian Cain, Mark Callahan, Shelley Callahan, Matt Christiansen, Lynn Coffman, Brian Crosby, Scott Darrington, Adam Eisenbarth, Bill Gimple, Erick Hsieh, Todd Hunt, Levi Kinateder, Colton Kohler, Joanna Kuzia, Les Kuzia, Ethan Lopes, Ian Mahaffey, Jason Matkins, Andrew Maynesik, Terry Miller, Seth Nuti, Melanie Phillips, Jiawei Simon Qin, Jim Reichle, Dan Ruby, Jeff Schloetter, David Schulz, Jeannie Smith, Kathy Trujillo, Jacob Wagner, Andrew Yoder, Ted Zel, and all people involved in this occultation observation.

The RECON project would not be possible without all of the support from
our community teams (teachers, students, and community members) and was
funded by NSF grants AST-1212159 and AST-1413287. Special thanks to Dean and Starizona for their support of RECON above and beyond the usual bounds of a commercial relationship.

Part of the research leading to these results has received funding from the European Research Council under the European Community's H2020 (2014-2020/ ERC Grant Agreement n° 669416 ``LUCKY STAR'').

Funding from Spanish grant AYA-2014-56637-C2-1-P is acknowledged, as is the Proyecto de Excelencia de la Junta de Andalucía, J. A. 2012-FQM1776. R.D. acknowledges the support of MINECO for his Ramon y Cajal Contract. FEDER funds are also acknowledged.

The authors want to thank the partial use of OPD/LNA facilities for this work.

ADO is thankful for the support of the CAPES (BEX 9110/12-7) FAPERJ/PAPDRJ (E-26/200.464/2015) grants.

ARGJ thanks CAPES/Brazil.

GBR is thankful for the support of CAPES/Brazil and FAPERJ (Grant E-01/2015/209545).

JIBC acknowledges CNPq for a PQ2 fellowship (process number 308489/2013-6).

LG thanks the support from CONACYT through grant 167236.

MA thanks the CNPq (Grants 473002/2013-2 and 308721/2011-0) and FAPERJ (Grant E-26/111.488/2013). 

PS-S acknowledge that part of the research leading to these results has received funding from the European Union’s Horizon 2020 Research and Innovation Programme, under Grant Agreement no 687378.

RVM acknowledges the following grants: CNPq-306885/2013, CAPES/Cofecub-2506/2015, FAPERJ/PAPDRJ-45/2013, FAPERJ/CNE/05-2015.

This research was partially based on data obtained at the Lowell Observatory's Discovery Channel Telescope (DCT). Lowell operates the DCT in partnership with Boston University, Northern Arizona University, the University of Maryland, and the University of Toledo. Partial support of the DCT was provided by Discovery Communications. LMI was built by Lowell Observatory using funds from the National Science Foundation (AST-1005313). A. Thirouin, S. Levine, and B. Skiff acknowledge Lowell Observatory funding.

The MEarth project acknowledges funding from the National Science Foundation, the David and Lucile Packard Foundation Fellowship for Science and Engineering, and the the John Templeton Foundation. The opinions expressed here are those of the authors and do not necessarily reflect the views of the John Templeton Foundation.

\clearpage

\clearpage
\begin{deluxetable}{cccccc}
\tabletypesize{\scriptsize}
\tablecaption{\label{tab:SitesPositive}Circumstances of observation for stations that acquired data.}
\tablewidth{0pt}
\tablehead{
 & \colhead{Longitude (W)} & \colhead{Telescope$^{a}$} & \colhead{Exposure} & Light curve & \colhead{Observer} \\
\colhead{Site} & \colhead{Latitude (N)} & \colhead{Camera} & \colhead{Time} & RMS & \\
& \colhead{Altitude (m)} &  & (s) & GBR/MWB & \colhead{Note}
}
\startdata
 & 119$^{\circ}$ 45' 53.0'' & Standard RECON & 2 & 0.235 / 0.140 & Dan Ruby, Brian Crosby, \\
Reno & 39$^{\circ}$ 23' 28.5'' & hardware setup$^{b}$ &  &  & Seth Nuti \\
 & 1470 & &  &  & (RECON) \\
\hline
Jack C. Davis Observatory & 119$^{\circ}$ 47' 46.8'' & Meade LX-200 & 2 & 0.212 / 0.194 & Jim Bean, Ethan Lopes \\
 & 39$^{\circ}$ 11' 08.2'' & 30 cm telescope &  &  & \\
`Carson City (B)' & 1548.1 & MallinCAM B\&W 428 &  &  & (RECON) \\
\hline 
 & 119$^{\circ}$ 33' 31.4'' & Standard RECON & 2 & 0.191 / 0.123 & Red Sumner \\
Carson City (S) & 39$^{\circ}$ 16' 26.5'' & hardware setup &  &  & \\
 & 1332.6 &  &  &  & (RECON) \\
\hline 
 & 119$^{\circ}$ 40' 20.3'' & Meade LX-200 & 2 & 0.197 / 0.123 & Jerry Bardecker \\
Gardnerville & 38$^{\circ}$ 53' 23.5'' & 30cm telescope &  &  & \\
 & 1534.9 & MallinCAM B\&W 428 &  &  & (RECON) \\
\hline 
 & 119$^{\circ}$ 09' 39.0'' & Standard RECON & 2 & 0.205 / 0.131 & Todd Hunt, Scott Darrington, \\
Yerington & 38$^{\circ}$ 59' 28.3'' & hardware setup &  &  & Joanna Kuzia, Les Kuzia, \\
 & 1342.7 &  &  &  & Matt Christiansen \\
 &  &  &  &  & (RECON) \\
\hline
 & 117$^{\circ}$ 14' 06.7'' & Standard RECON & 2 & 0.214 / 0.128 & Teralyn Blackburn, \\
Tonopah & 38$^{\circ}$ 05' 22.1'' & hardware setup &  &  & Clair Blackburn \\
 & 1838.7 &  &  &  & (RECON) \\
\hline 
 & 088$^{\circ}$ 11' 46.4'' & 50 cm Newtonian & 4 & 0.180 / 0.168 & Aart Olsen \\
Urbana & 40$^{\circ}$ 05' 12.5'' & Watec 120N+ &  &  & \\
 & 227 &  &  &  & \\
\hline
 & 115$^{\circ}$ 27' 58.0'' & OAN/SPM Harold  & 2 & - & Leonel Gutierrez et al. \\
San Pedro Martir & 31$^{\circ}$ 02' 42.0'' & L. Johnson 1.5 m  telescope &  &  & \\
 & 2790 & FLI ProLine &  &  & \\
 &  & PL3041 (PL0212309)  &  &  & \\
\hline 
San Pedro Martir & 115$^{\circ}$ 28' 00.0'' & 0.84 m telescope & 5 & - & Marco G\'{o}mez et al. \\
 & 31$^{\circ}$ 02' 43.0'' & SPECTRAL E2V-4240 &  &  & \\
 & 2790 & Mexman &  &  & \\
\enddata

\tablenotetext{a}{For more details on the RECON equipment, see \citet{Buie2016RECON}.}
\tablenotetext{b}{Standard RECON hardware setup consist in a 28 cm Celestron CPC1100 telescope and a MallinCAM B\&W 428.}

\end{deluxetable}

\clearpage
\begin{deluxetable}{ccccc}
\tabletypesize{\scriptsize}
\tablecaption{\label{tab:SitesNegative}Circumstances of observation for stations with no data acquired.}
\tablewidth{0pt}
\tablehead{
 & \colhead{Longitude (W)} & \colhead{Telescope$^{a}$} & \colhead{Result} & \colhead{Observer} \\
\colhead{Site} & \colhead{Latitude (N)} & \colhead{Camera} &  &  \\
& \colhead{Altitude (m)} &  &  & \colhead{Note}
}
\startdata
Lowell Observatory & 111$^{\circ}$ 32' 09.0'' & 1.1 m Hall & Clouds & Brian Skiff \\
Anderson Mesa & 35$^{\circ}$ 05' 49.0'' & nasa42 &  & \\
 & 2163 &  &  & \\
\hline
 & 120$^{\circ}$ 09' 9.4'' & Standard RECON & Clouds & Brian Cain, Terry Miller, \\
Cedarville & 41$^{\circ}$ 31' 50.0'' & hardware setup &  & David Schulz \\
 & 1381.4 &  &  & (RECON) \\
\hline
 & 116$^{\circ}$ 42' 48.9'' & Meade LX-200 & Clouds and & John Keller, Melanie Phillips,  \\
CPSLO & 33$^{\circ}$ 44' 3.4'' & 30 cm telescope & Focus & Eric Hsieh, Ian Mahaffey, \\
Idyllwild/Astrocamp & 1714.2 & MallinCAM B\&W 428 & problems & Tedd Zel, Jeff Schloetter, \\
 & & & & Andrew Yoder, Jiawei Simon Qin \\
 & & & & Jacob Wagner, Adam Eisenbarth \\
 & & & & (RECON) \\
\hline
Lowell Observatory & 111$^{\circ}$ 25' 20.0'' & 4.3m & Clouds & Stephen Levine \\
Discovery Channel & 34$^{\circ}$ 44' 40.0'' & Large Monolith Imager &  & \\
Telescope & 2360 &  &  & \\
\hline
 & 121$^{\circ}$ 23' 56.'' &  Standard RECON & Clouds & Andrew Mayncsik \\
Fall River/Burney & 41$^{\circ}$ 02' 45.''  & hardware setup &  & \\
 & 1012 &  &  & (RECON) \\
\hline
Fred Lawrence & 110$^{\circ}$ 52' 42.0'' & MEarth & Clouds & J. Irwin, \\
Whipple Observatory & 31$^{\circ}$ 40' 52.0'' &  &  & D. Charbonneau \\
 & 2606 &  &  & \\
\hline
 & 120$^{\circ}$ 57' 04'' & Standard RECON & Clouds & Bill Gimple, Barclay Anderson \\
Greenville & 40$^{\circ}$ 08' 23.'' & hardware setup &  & \\
 & 1098 &  &  & (RECON) \\
\hline
 & 118$^{\circ}$ 37' 49'' & Standard RECON & Instrument & Kathy Trujillo \\
Hawthorne & 38$^{\circ}$ 31' 35'' & hardware setup & problems &  \\
 & 1321 &   &  & (RECON) \\
\hline
 & 120$^{\circ}$ 34' 03.5'' & Standard RECON & Frost/Clouds &  Shelley Callahan, \\
Portola & 39$^{\circ}$ 43' 56.1'' & hardware setup &  & Mark Callahan \\
 & 1351 &  &  & (RECON) \\
\hline
 & 120$^{\circ}$ 58' 10.3'' & Standard RECON & Clouds & Charley Arrowsmith, Lynn Coffman, \\
Quincy & 39$^{\circ}$ 56' 52.7'' & hardware setup &  & Levi Kinateder, Colton Kohler, \\
Feather River College & 1046 &  &  & Mystery Brown \\
 & & & & (RECON) \\
\hline
 & 119$^{\circ}$ 45' 53.0'' & Meade LX-200 & Clouds & Buck Bateson \\
Susanville & 39$^{\circ}$ 23' 28.4'' & 30 cm telescope &  & \\
 & 1470 & MallinCAM B\&W 428 &  & (RECON) \\
\hline
 & 121$^{\circ}$ 28' 44.'' & Standard RECON & Clouds & Jason Matkins, Jeannie Smith \\
Tulelake & 41$^{\circ}$ 57' 19.'' & hardware setup &  & \\
 & 1232 &  &  & (RECON) \\
\enddata

\tablenotetext{a}{For more details on the RECON equipment, see \citet{Buie2016RECON}}
\end{deluxetable}

\begin{center}
\begin{figure}[H]
\begin{centering}
\epsscale{.8}
\plotone{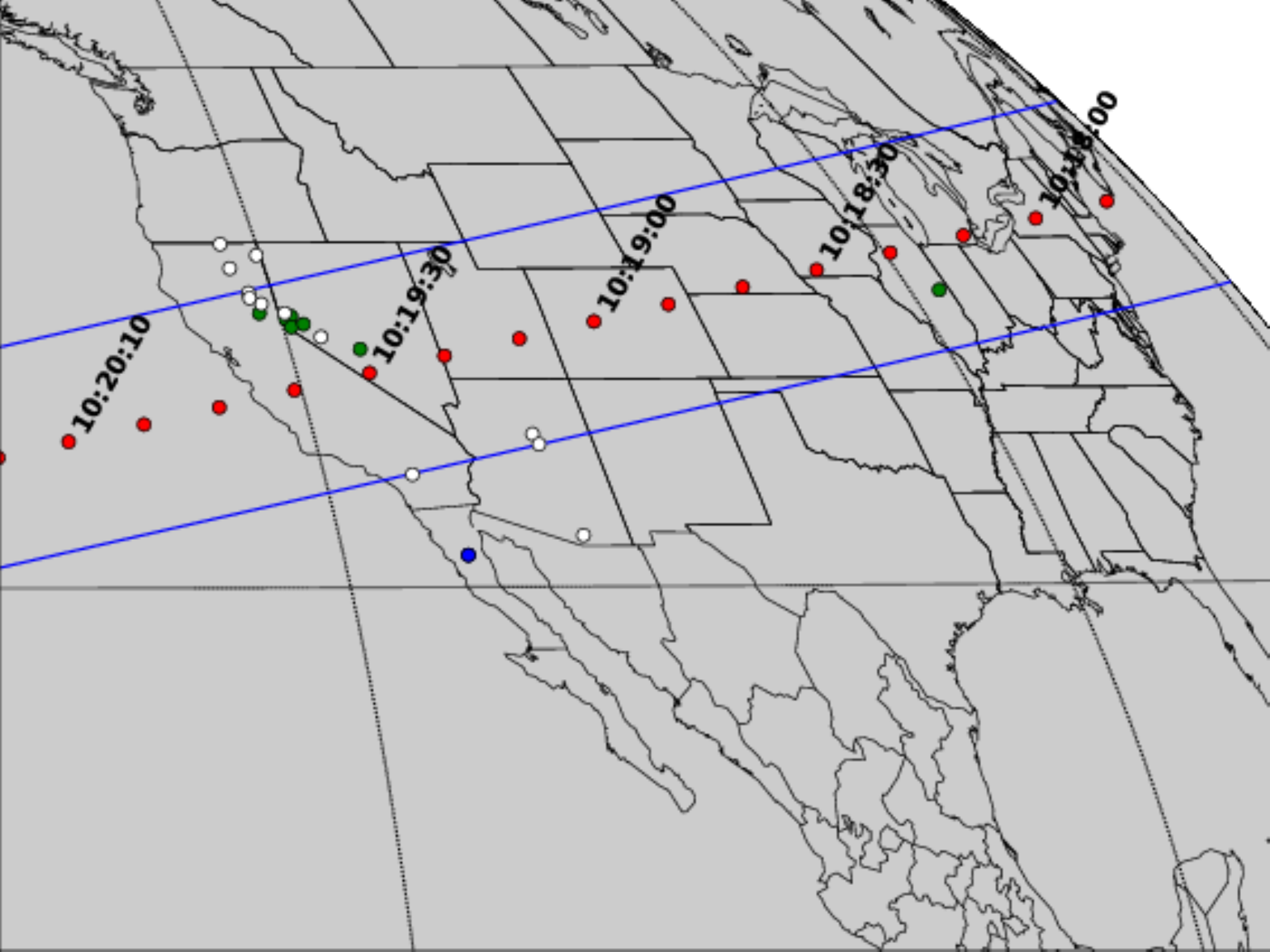}
\par\end{centering}

\caption{\label{fig:PredicMap}Post-occultation reconstruction of the 2007 UK$_{126}$'s
shadow path on Earth for the 2014 November 15 event. The shadow moves
from right to left; blue lines are the expected size limit of the TNO (km), as derived from Figs. \ref{fig:NominalSolutionGBR} and \ref{fig:NominalSolutionMWB};
the red dots represents the center of the body for a given time, each separated by 10 seconds. The green dots are the sites where the occultation was
detected (Table \ref{tab:SitesPositive}). The blue dots are the two telescopes at San Pedro Martir, that acquired data but did not detect the event, and the white dots are the sites that were clouded out or suffered technical failures (Table \ref{tab:SitesNegative}).}
\end{figure}
\par\end{center}

\begin{center}
\begin{figure}[H]
\begin{centering}
\includegraphics[scale=0.5]{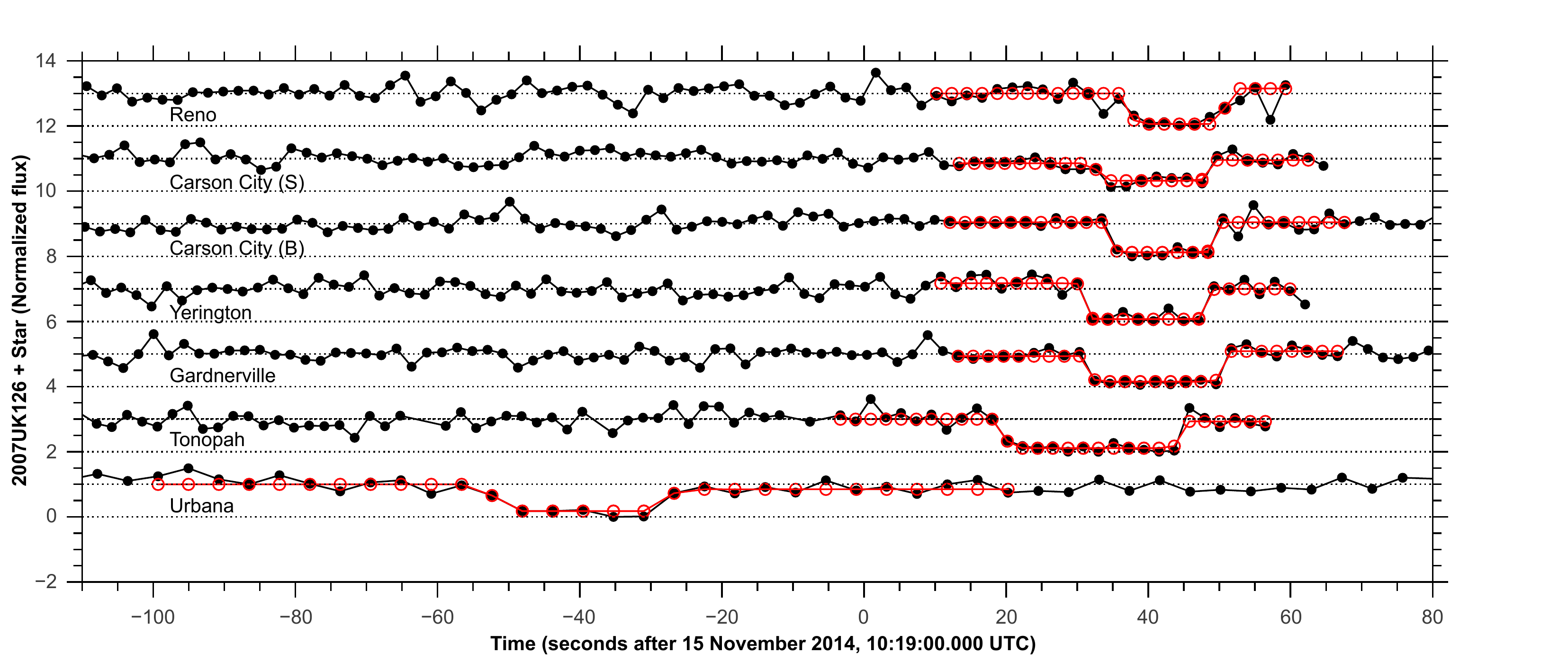}
\par\end{centering}

\begin{centering}
\includegraphics[scale=0.5]{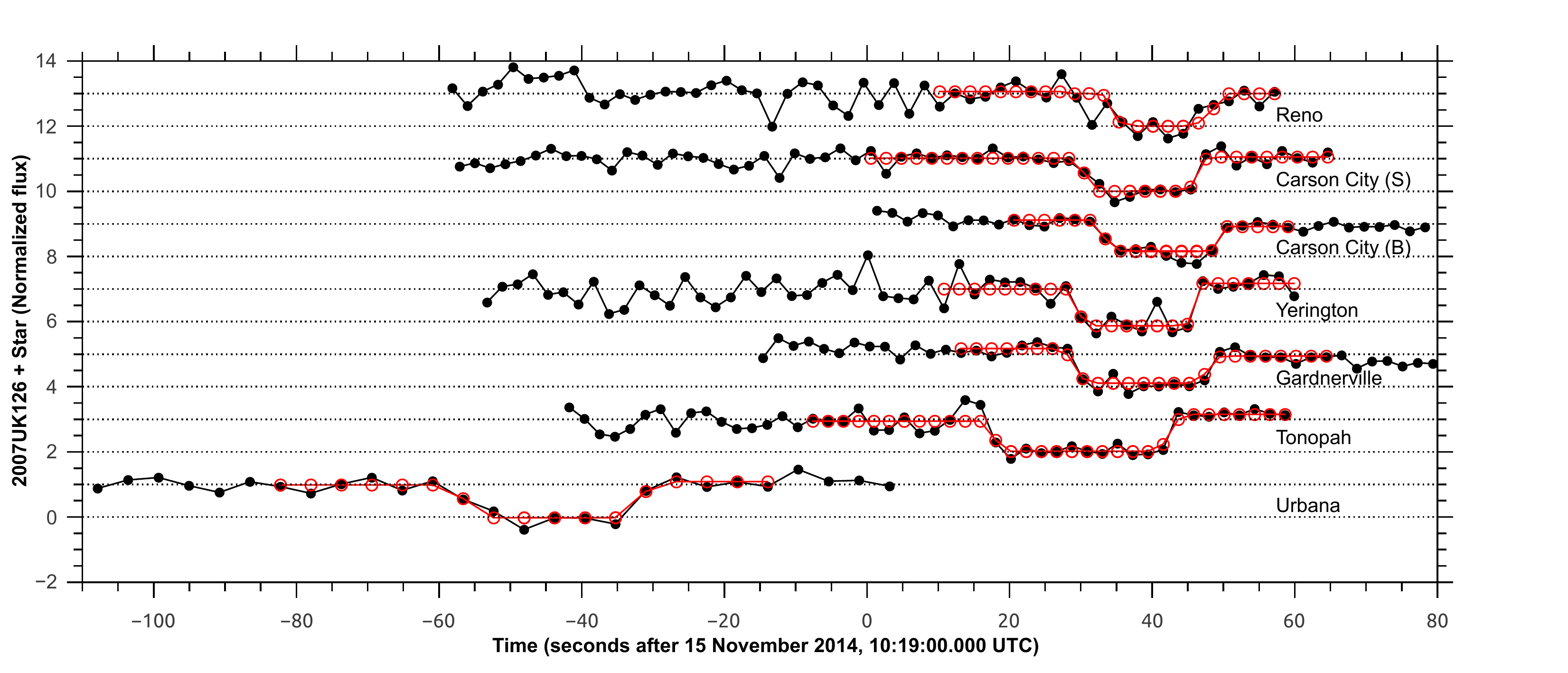}
\par\end{centering}

\caption{\label{fig:LightCurves}Top panel: the seven occultation light curves, obtained from the GBR extraction (see text), normalized to the unocculted star plus 2007 UK$_{126}$ flux and vertically shifted by integer values for better viewing. The red lines and
circles are the square-well model convoluted by the Fresnel diffraction,
the star apparent angular diameter, and the finite exposure time. The mid-times of the occultations do not coincide due to the propagation delays of
the shadow due to the distinct longitudes of the sites (Fig. \ref{fig:PredicMap}).
Bottom panel: Same from the MWB extraction. Note that no secondary occultation is observed, as could be caused by a satellite.}
\end{figure}
\par\end{center}

\clearpage
\begin{deluxetable}{cccccc}
\tabletypesize{\scriptsize}
\tablecaption{\label{tab:Instant_Ingress_egress}Disappearance (ingress) and
re-appearance (egress) times.}
\tablewidth{0pt}
\tablehead{
\colhead{Site} & \colhead{GBR} & \colhead{Error} & \colhead{MWB} & \colhead{Error} & \colhead{Difference$^{a}$} \\
& \colhead{Ingress (UTC)} & \colhead{(s)} & \colhead{Ingress (UTC)} & \colhead{(s)} & \colhead{(s)} \\
& \colhead{Egress (UTC)} & \colhead{} & \colhead{Egress (UTC)} & \colhead{} &
}

\startdata
\rule{0pt}{15pt}\multirow{2}{*}{Reno} & 10:19:35.02 & 0.80 & 10:19:34.10 & 0.70 & 0.92 \\
 & 10:19:46.68 & 0.47  & 10:19:46.64 & 0.65 & 0.04 \\
\multirow{2}{*}{Carson City (S)} & 10:19:30.85 & 0.66 & 10:19:30.60 & 0.60 & 0.25 \\
 & 10:19:46.32 & 0.41 & 10:19:46.30 & 0.42 & 0.02 \\
\multirow{2}{*}{Carson City (B)} & 10:19:32.47 & 0.51 & 10:19:32.20 & 0.50 & 0.27 \\
 & 10:19:47.19 & 0.35 & 10:19:47.30 & 0.41 & -0.11 \\
\multirow{2}{*}{Yerington} & 10:19:29.42 & 0.40 & 10:19:29.25 & 0.41 & 0.17 \\
 & 10:19:45.97 & 0.35 & 10:19:46.00 & 0.35 & -0.03 \\
\multirow{2}{*}{Gardnerville} & 10:19:29.40 & 0.42 & 10:19:29.71 & 0.46 & -0.31 \\
 & 10:19:48.35 & 0.29 & 10:19:48.07 & 0.30 & 0.28 \\
\multirow{2}{*}{Tonopah} & 10:19:17.49 & 0.54 & 10:19:17.60 & 0.55 & 0.11 \\
 & 10:19:42.39 & 0.26 & 10:19:42.45 & 0.30 & -0.06 \\
\multirow{2}{*}{Urbana} & 10:18:03.76 & 0.90 & 10:18:03.61 & 0.90 & 0.15 \\
 & 10:18:27.62 & 0.90 & 10:18:27.60 & 0.90 & 0.02 \\
\enddata

\tablenotetext{a}{Difference between times in the sense `GBR - MWB'.}
\end{deluxetable}

\begin{center}
\begin{figure}[H]
\begin{centering}
\includegraphics[scale=0.6]{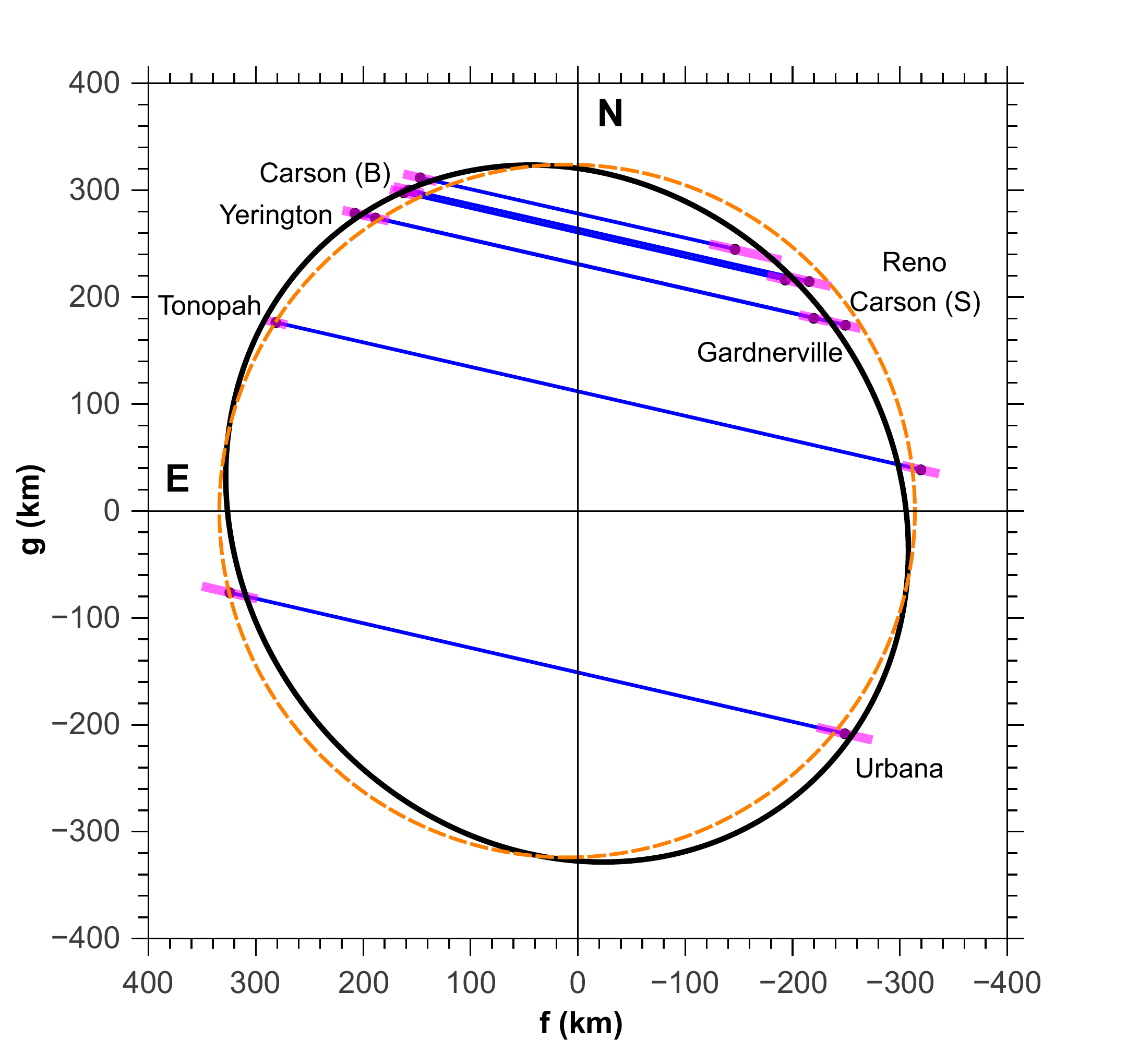}
\par\end{centering}

\caption{\label{fig:NominalSolutionGBR}The positive occultation chords and the GBR solution (see text). The best elliptical fit to the occultation chords are shown in black and the dashed orange line is the circular fit. Adjusted parameters are shown in Table \ref{tab:Physical-Parameters}. The magenta segments are the 1-$\sigma$ error bars on each occultation chord extremity.}
\end{figure}
\par\end{center}

\begin{center}
\begin{figure}[H]
\begin{centering}
\includegraphics[scale=0.6]{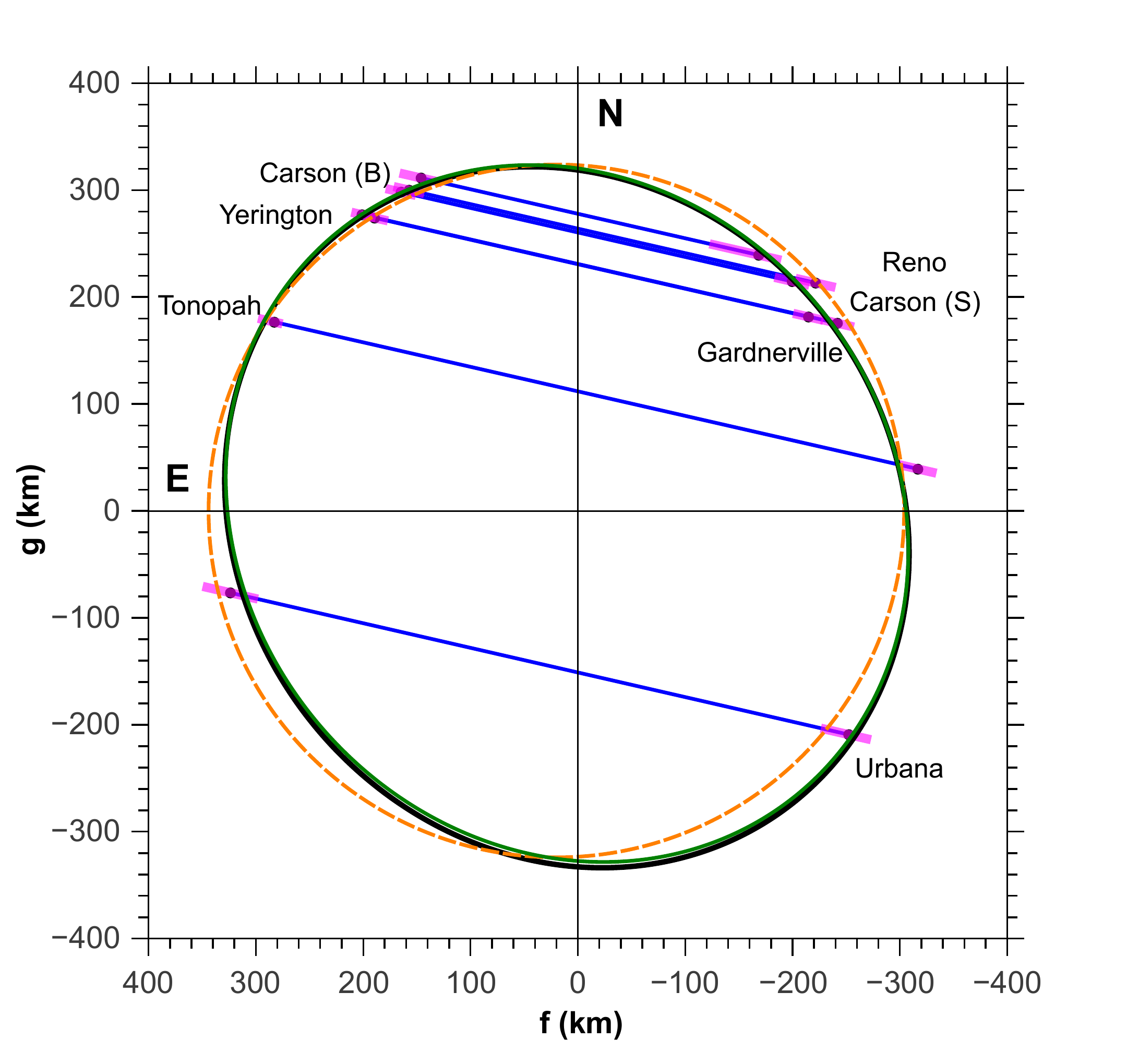}
\par\end{centering}

\caption{\label{fig:NominalSolutionMWB} The same as Fig. \ref{fig:NominalSolutionGBR} for the MWB solution. The best elliptical fit is in black and the fit from GBR is shown in green. Note that the two solutions are very close to each other, with a maximum radial discrepancy of about 5 km.}
\end{figure}
\par\end{center}

\clearpage
\begin{deluxetable}{cccc}
\tabletypesize{\scriptsize}
\tablecaption{\label{tab:Physical-Parameters}Physical Parameters of 2007 $UK_{126}$
from the two studied solutions.}
\tablewidth{0pt}
\tablehead{
\colhead{Solution} & \colhead{GBR} & \colhead{MWB} 
}
\startdata
Semimajor axis (km) & 339$_{-10}^{+15}$ & $340_{-8}^{+12}$ \\
Equivalent Radius (km) & $319_{-7}^{+14}$ & $319_{-6}^{+12}$ \\
Circular fit Radius (km) &  $324_{-23}^{+27}$ & $328_{-21}^{+26}$ \\
Apparent oblateness & 0.106$_{-0.040}^{+0.050}$ & $0.118 _{-0.048}^{+0.055}$  \\
$f{}_{c}$ (km) & $-3699\pm12$ & $-3699\pm13$  \\
$g{}_{c}$ (km) & $-3457\pm13$ & $-3456\pm13$  \\
Position angle (deg) & $129_{-22}^{+14}$ & $134_{-17}^{+14}$  \\
$\chi_{pdf}^{2}$ (elliptical fit) & 0.71 & 0.65  \\
$\chi_{pdf}^{2}$ (circular fit) & 1.65 & 1.45  \\
Radial rms (km) & 10.6 & 10.9 \\
$p_{R}$ (Thirouin)$^{a}$ & $0.189_{-0.015}^{+0.009}$ & $0.189_{-0.013}^{+0.008}$  \\
$p_{V}$ (Perna)$^{b}$ & $0.159_{-0.013}^{+0.007}$ & $0.159_{-0.011}^{+0.006}$  \\
Density (kg~m$^{-3}$)$^{c}$ & $<1740$ & $<1620$  \\
\enddata 

\tablenotetext{a}{`$p_{R}$ (Thirouin)' means $p_{R}$ using $H$
from MPC and \citet{2014A&A...569A...3T}}
\tablenotetext{b}{`$p_{V}$ (Perna)' means $p_{V}$ using $H$ from \citet{2010A&A...510A..53P}}
\tablenotetext{c}{Upper limit based on a hydrostatic shape with a rotation period of 11.05h}
\end{deluxetable}

\begin{center}
\begin{figure}[H]
\begin{centering}
\includegraphics[scale=0.7]{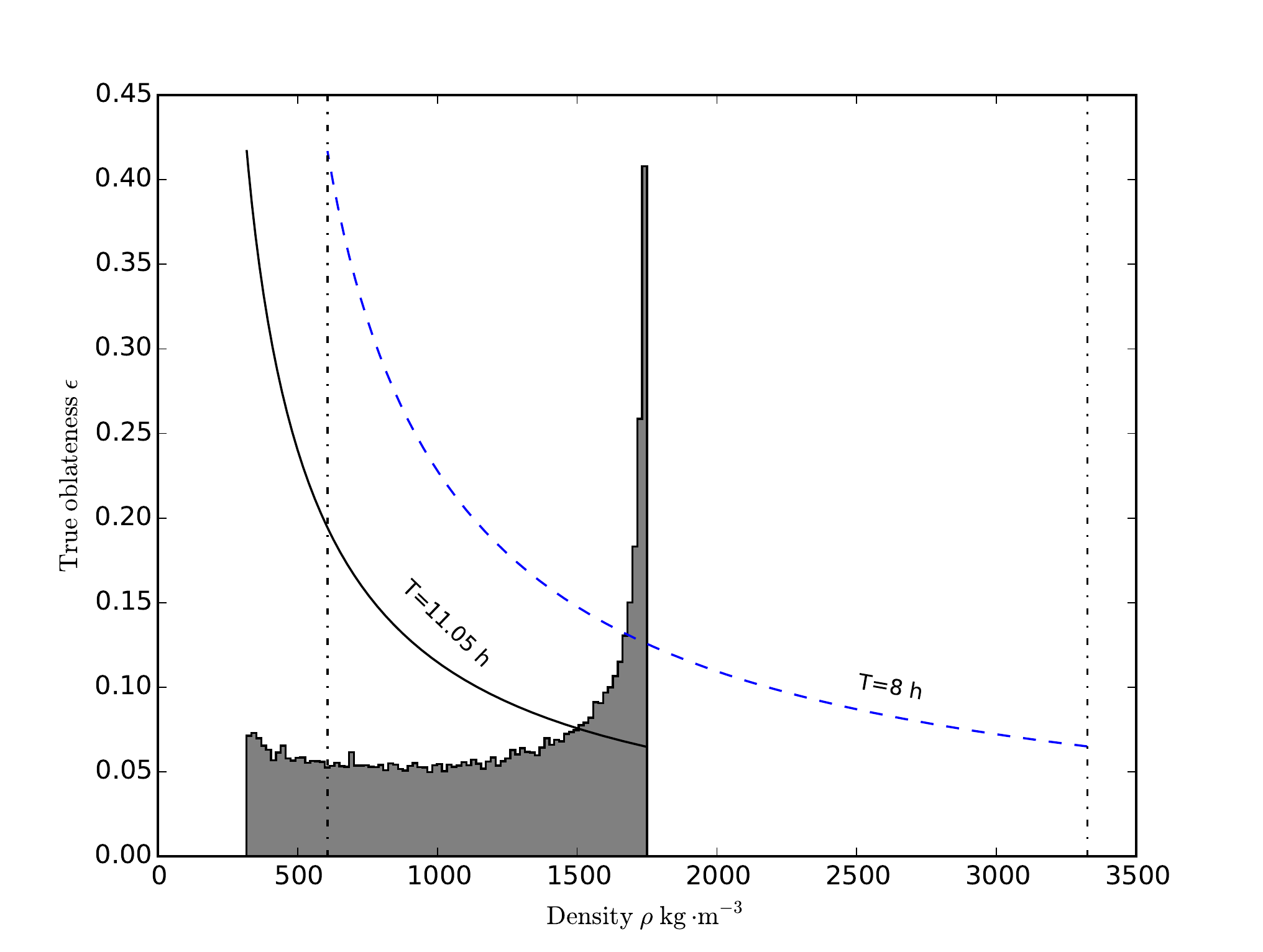}

\par\end{centering}

\caption{\label{fig:maclaurin_and_density} The continuous line is the Maclaurin equilibrium curve for the preferred period of 11.05 hours. The lower limit is given by the stability condition for a Maclaurin spheroid, while the upper limit is given by the occultation. Gray-filled histogram is the probability distribution for the density within this range in arbitrary units. Dashed blue line is the oblateness density relation for a period T=8 h, the lowest probable rotation period (see text). In this case, the probability distribution is qualitatively the same (not plotted) but in the range between 600 and 3300 kg~m$^{-3}$ indicated by vertical dash-dotted lines.}
\end{figure}
\par\end{center}

\end{document}